\documentclass[12pt]{article}
\usepackage{fullpage,enumitem,amsmath,amssymb,graphicx}
\usepackage{subcaption}
\usepackage{caption}
\usepackage{braket}
\usepackage{authblk}
\usepackage{tikz}
\usepackage{quantikz}
\usepackage{multicol}
\usepackage{float}

\usepackage[
backend=biber,
style=numeric
]{biblatex}

\usepackage{hyperref} 
\hypersetup{
    colorlinks=true,
    linkcolor=blue,
    filecolor=magenta,      
    urlcolor=cyan,
}

\title{
    \textbf{Enhancing Quantum Diffusion Models for Complex Image Generation} \\        
    \vspace{0.5em}              
}

\author[1]{Jeongbin Jo}
\author[2]{Santanam Wishal}
\author[3]{Shah Md Khalil Ullah}
\author[4]{Shan Zeng}
\author[5]{Dikshant Dulal}

\affil[1]{Department of Physics \\ Yonsei University \\ \textit{jeongbin033@yonsei.ac.kr}}
\affil[2]{Asia Cyber University \\ \textit{santawishal17@gmail.com}}
\affil[3]{Khulna University of Engineering and Technology \\ \textit{hamimkhandakar222@gmail.com}}
\affil[4]{Eleven Dimensions LLC \\ \textit{shanjdk2012@gmail.com}}

\begin{document}
\maketitle
\begin{abstract}
Quantum generative models offer a novel approach to exploring high-dimensional Hilbert spaces but face significant challenges in scalability and expressibility, particularly when applied to multi-modal distributions.
In this study, we propose a \textbf{Hybrid Quantum-Classical U-Net} architecture enhanced by \textbf{Adaptive Non-Local Observables (ANO)} and an \textbf{Ancilla-based Global Feature Extractor}. 
By compressing classical data into a dense quantum latent space and utilizing trainable observables, our model extracts rich non-local features that complement classical processing. 
Furthermore, we integrate a Hadamard Test module to capture global structural information, fusing it with dense local features. 
We also investigate the role of Skip Connections in preserving semantic information during the reverse diffusion process. 
Experimental results on the full MNIST dataset (digits 0-9) demonstrate that the proposed architecture generates structurally coherent and recognizable images for all digit classes, overcoming the mode collapse observed in prior quantum diffusion models. 
While hardware constraints necessitate resolution downscaling, our findings suggest that hybrid architectures with adaptive measurements provide a feasible pathway for enhancing generative capabilities in the NISQ era.
\end{abstract}

\tableofcontents
\begin{center}

\end{center}

\noindent
\rule{\linewidth}{0.4pt}

  
\section{Introduction}
Quantum Machine Learning (QML) is an emerging interdisciplinary research field that integrates principles of quantum computing with classical machine learning techniques to enhance data processing, pattern recognition, and optimization tasks. 
By exploiting quantum mechanical phenomena such as superposition, entanglement, and quantum interference, QML aims to provide computational advantages over classical learning algorithms, particularly for high-dimensional and complex datasets.

In classical machine learning, the performance of algorithms is often constrained by computational complexity and memory requirements as data size and model depth increase. 
Quantum computing, on the other hand, offers a fundamentally different computation paradigm, where information is represented using quantum bits (qubits) that can exist in multiple states simultaneously. 
This property allows quantum algorithms to explore large solution spaces more efficiently than their classical counterparts in certain scenarios.

One of the key motivations behind QML is the potential for speedup in tasks such as linear algebra operations, optimization, and sampling, which form the backbone of many machine learning models. 
Quantum-enhanced algorithms, including quantum support vector machines, variational quantum circuits, and quantum neural networks, have been proposed to accelerate learning processes and improve model expressiveness.

Despite its promising advantages, QML also faces several challenges. 
Current quantum hardware, commonly referred to as Noisy Intermediate-Scale Quantum (NISQ) devices, suffers from limited qubit counts, short coherence times, and gate errors. 
These hardware constraints restrict the depth and scalability of quantum models, often requiring hybrid quantum-classical approaches for practical implementations. 
Additionally, the theoretical quantum advantage of many QML algorithms remains an open research question, as classical algorithms continue to improve rapidly.

In summary, Quantum Machine Learning represents a promising yet evolving research direction. 
While it offers the potential for enhanced computational power and novel learning capabilities, significant challenges related to hardware limitations, noise, and algorithmic scalability must be addressed before widespread real-world adoption becomes feasible.

\section{Diffusion Model}
  \subsection{Classical Diffusion Model}
  Diffusion models are a class of generative models in machine learning inspired by non-equilibrium thermodynamics. The goal is to learn a diffusion process that destroys structure in data, and then learn a reverse process that restores structure to generate new data samples from noise.\cite{diffusion-model}

    \subsubsection{Forward Process: Diffusion Process}
    The forward process is defined by a Markov chain that gradually adds Gaussian noise to the data according to a variance schedule $\beta_1, \dots, \beta_T$.
    \begin{equation}
      q(x_t | x_{t-1}) = \mathcal{N}\left(x_t ; \sqrt{1-\beta_t}x_{t-1}, \beta_t \mathbf{I}\right)
    \end{equation}

    Using the notation $\alpha_t = 1 - \beta_t$ and $\bar{\alpha}_t = \prod_{s=1}^t \alpha_s$, we can sample $x_t$ at any arbitrary time step $t$ directly from $x_0$ in closed form:
    \begin{equation}
      q(x_t | x_0) = \mathcal{N}\left(x_t ; \sqrt{\bar{\alpha}_t}x_0, (1 - \bar{\alpha}_t)\mathbf{I}\right)
    \end{equation}
    This property allows for efficient training by eliminating the need to iterate through all previous timesteps.

    \subsubsection{Reverse Process: Generative Process}
    The generative process\cite{DBLP:journals/corr/abs-2006-11239} is defined as the reverse Markov chain. 
    Since the exact reverse posterior $q(x_{t-1}|x_t)$ is intractable, we approximate it using a neural network with parameters $\theta$:
    \begin{equation}
      p_\theta(x_{t-1} | x_t) = \mathcal{N}\left(x_{t-1} ; \mu_\theta(x_t, t), \Sigma_\theta(x_t, t) \right)
    \end{equation}
    Here, $\mu_\theta(x_t, t)$ is the predicted mean, and $\Sigma_\theta(x_t, t)$ is the covariance (often fixed to $\beta_t \mathbf{I}$ or learned).

    \subsubsection{Loss Function}
    Training is performed by optimizing the variational lower bound (VLB) on the negative log-likelihood.
    \begin{equation}
      \mathcal{L} = \mathbb{E} \left[ - \log p_\theta(\mathbf{x}_0) \right] \leq \mathbb{E}_q \left[ - \log \frac{p_\theta(\mathbf{x}_{0:T})}{q(\mathbf{x}_{1:T} | \mathbf{x}_0)} \right]
    \end{equation}

    This can be rewritten as a sum of KL divergence\cite{kl-divergence} terms:
    \begin{equation}
      \mathcal{L} = \mathbb{E}_q \left[ \underbrace{D_{KL}(q(\mathbf{x}_T|\mathbf{x}_0) || p(\mathbf{x}_T))}_{L_T} + \sum_{t>1} \underbrace{D_{KL}(q(\mathbf{x}_{t-1}|\mathbf{x}_t, \mathbf{x}_0) || p_\theta(\mathbf{x}_{t-1}|\mathbf{x}_t))}_{L_{t-1}} \underbrace{- \log p_\theta(\mathbf{x}_0|\mathbf{x}_1)}_{L_0} \right]
    \end{equation}

    In practice, Ho et al. found that a simplified objective function yields better sample quality. The objective is to predict the noise $\epsilon$ added to $x_0$:
    \begin{equation}
      \mathcal{L}_{\text{simple}}(\theta) = \mathbb{E}_{t, x_0, \epsilon} \left[ \| \epsilon - \epsilon_\theta(\sqrt{\bar{\alpha}_t}x_0 + \sqrt{1-\bar{\alpha}_t}\epsilon, t) \|^2 \right]
    \end{equation}
    where $\epsilon \sim \mathcal{N}(0, \mathbf{I})$ and $\epsilon_\theta$ is a function approximator (e.g., U-Net).

    \subsubsection{Continuous-Time Formulation: SDE}
    The discrete diffusion process can be generalized to continuous time using Stochastic Differential Equations (SDEs). 
    The forward process is described by the following It\^{o} SDE\cite{Ito-calculus}:
    \begin{equation}
      d\mathbf{x} = \mathbf{f}(\mathbf{x}, t)dt + g(t)d\mathbf{w}
    \end{equation}
    where $\mathbf{f}(\cdot, t)$ is the drift coefficient, $g(t)$ is the diffusion coefficient, and $\mathbf{w}$ is a standard Wiener process.

    Remarkably, the reverse process—generating data from noise—is also a diffusion process governed by the \textit{reverse-time SDE}\cite{ANDERSON1982313}:
    \begin{equation}
      d\mathbf{x} = \left[ \mathbf{f}(\mathbf{x}, t) - g(t)^2 \nabla_\mathbf{x} \log p_t(\mathbf{x}) \right] dt + g(t) d\bar{\mathbf{w}}
    \end{equation}
    Here, $dt$ represents a negative time step, and $\bar{\mathbf{w}}$ is the Brownian motion in reverse time. 
    The term $\nabla_\mathbf{x} \log p_t(\mathbf{x})$, known as the \textit{score function}, points towards high-density regions of the data.

    Therefore, the core task of the diffusion model is to learn a score-based model 
    $s_\theta(\mathbf{x}, t) \approx \nabla_\mathbf{x} \log p_t(\mathbf{x})$ using a neural network (or PQC in our case), 
    allowing us to numerically solve the reverse SDE to generate samples from noise.

  \subsection{Quantum Diffusion Model}
  While classical diffusion models operate on probability distributions over classical data, 
  Quantum Diffusion Models (QDMs) extend this concept to the Hilbert space of quantum states. 
  The goal is to generate quantum states (density matrices) from a maximally mixed state.

    \subsubsection{Forward Process: Depolarizing Channel}
    Instead of adding Gaussian noise, the forward process in QDM is typically modeled as a standard depolarizing channel acting on a density matrix $\rho$. 
    For a system of $n$ qubits with dimension $d=2^n$, the state at step $t$ is given by:

    \begin{equation}
      \rho_t = \mathcal{E}_t(\rho_{t-1}) = (1 - p_t)\rho_{t-1} + p_t \frac{I}{d}
    \end{equation}

    where $p_t \in [0, 1]$ is the depolarization probability (noise schedule). 
    Similar to the classical case, we can express $\rho_t$ directly from the initial state $\rho_0$. 
    Let $\alpha_t = \prod_{s=1}^t (1 - p_s)$, then:

    \begin{equation}
      \rho_t = \alpha_t \rho_0 + (1 - \alpha_t) \frac{I}{d}
    \end{equation}

    As $t \to T$, $\alpha_T \to 0$, and the state converges to the maximally mixed state $\rho_T \approx \frac{I}{d}$, 
    which contains no information about $\rho_0$.

    \subsubsection{Reverse Process: Quantum Denoising}
    The reverse process aims to restore the quantum state from the noise. 
    This is modeled by a parameterized quantum circuit (PQC), denoted as a unitary operator $U(\theta)$. 
    The discrete reverse step can be approximated as:

    \begin{equation}
      \rho_{t-1} \approx \mathcal{D}_\theta(\rho_t, t) = U(\theta_t) \rho_t U^\dagger(\theta_t)
    \end{equation}

    For more complex generative tasks, the reverse process may involve ancillary qubits and measurements to simulate non-unitary maps.

    \subsection{Loss Function: Quantum Infidelity Loss}
    Unlike classical diffusion models that typically minimize the Mean Squared Error (MSE) between predicted and actual noise, our Quantum U-Net operates directly on the Hilbert space. Therefore, we employ the infidelity Loss to maximize the overlap between the generated quantum state $\rho_\theta$ and the target state $\sigma$.

    The Fidelity $F$ between two pure states $\ket{\psi_\theta}$ and $\ket{\psi_{\text{target}}}$ is defined as $|\braket{\psi_{\text{target}}|\psi_\theta}|^2$. In our variational setting, we aim to minimize the infidelity:
    \begin{equation}
        \mathcal{L}(\theta) = 1 - F(\rho_\theta, \sigma) = 1 - \bra{\psi_{\text{target}}} \rho_\theta \ket{\psi_{\text{target}}}
    \end{equation}
    By minimizing this objective, the PQC learns to orient the qubit register such that the measured expectation values align with the target distribution, effectively reversing the diffusion process in the quantum feature space.

\section{Quantum Data Encoding}
Our encoding methods are based on IBM Quantum Platform, especially Quantum Machine Learning.\cite{Data-encoding}
  \subsection{Basis Encoding}
  Basis encoding maps a classical $P$-bit string directly to a computational basis state of a $P$-qubit system. 
  For a single feature represented as binary bits $(b_1, b_2, \dots, b_P)$, the quantum state is:
  \begin{equation}
    \ket{x} = \ket{b_1, b_2, \dots, b_P}, \quad b_i \in \{0, 1\}
  \end{equation}

  \subsection{Amplitude Encoding}
  Amplitude encoding maps a normalized classical $N$-dimensional data vector $\vec{x}$ to the amplitudes of an $n$-qubit quantum state, 
  where $n = \lceil \log_2 N \rceil$.

  \begin{equation}
    \ket{\psi_x} = \frac{1}{\alpha}\sum_{i=1}^N x_i \ket{i}
  \end{equation}
  Here, $\ket{i}$ is the computational basis state and $\alpha$ is a normalization constant ensuring $\langle \psi_x | \psi_x \rangle = 1$. 
  Here, $\alpha$ is a normalization constant to be determined from the data being encoded. We use amplitude encoding in the quantum diffusion model.

  \begin{equation}
    \alpha = \sqrt{\sum_{i=1}^{N}|x_i|^2}
  \end{equation}

  \subsection{Angle Encoding}
  Angle encoding maps each feature $x_k$ to the rotation angle of a qubit using $R_Y$ gates. 
  For a data vector $\vec{x}$, the state is a product state:
  \begin{equation}
    \ket{\vec{x}} = \bigotimes_{k=1}^N R_Y(x_k)\ket{0} = 
    \bigotimes_{k=1}^N \left( \cos\left(\frac{x_k}{2}\right)\ket{0} + \sin\left(\frac{x_k}{2}\right)\ket{1} \right)
  \end{equation} 
  \subsection{Phase Encoding}
  Phase encoding maps data features to the phase of qubits using Phase gates $P(\phi)$, typically applied after Hadamard gates.
  \begin{equation}
    \ket{\vec{x}} = \bigotimes_{k=1}^N P(x_k)\ket{+} = 
    \frac{1}{\sqrt{2^N}} \bigotimes_{k=1}^N \left( \ket{0} + e^{i x_k}\ket{1} \right)  
  \end{equation} 

  \subsection{Dense Angle Encoding}
  Dense angle encoding encodes two features, $x_k$ and $x_l$, into a single qubit using both $Y$-axis and $Z$-axis rotations.
  \begin{equation}
    \ket{x_k, x_l} = R_Z(x_l) R_Y(x_k)\ket{0} = \cos\left(\frac{x_k}{2}\right)\ket{0} + e^{i x_l} \sin\left(\frac{x_k}{2}\right)\ket{1} 
  \end{equation}

  Extending this to more features, the data vector $\vec{x} = (x_1, ..., x_N)$ can be encoded as:
  \begin{equation}
    \ket{\vec{x}} = \bigotimes_{k=1}^{N/2}\left( \cos{x_{2k-1}}\ket{0} + e^{ix_{2k}}\sin{x_{2k-1}}\ket{1} \right)
  \end{equation}

\section{Quantum Circuit}
The choice of quantum circuit ansatz is critical in designing Quantum Machine Learning (QML) models, 
as it determines the expressivity and trainability of the network within the constraints of Noisy Intermediate-Scale Quantum (NISQ) devices. 
Based on recent surveys and specific applications in diffusion models, we categorize relevant ansatzes into general-purpose architectures and task-specific designs for image processing.

\subsection{General Variational Ansatzes}
\begin{itemize}
    \item \textbf{Hardware-Efficient Ansatz (HEA):} 
    Designed to minimize circuit depth and gate count on NISQ devices, HEA utilizes a layered structure of parameterized single-qubit rotations followed by entangling gates (e.g., CNOTs) tailored to the specific connectivity of the quantum hardware \cite{guo2024patterns}. 
    While it offers high implementability, it is known to suffer from barren plateaus if not carefully initialized.
    
    \item \textbf{Hamiltonian Variational Ansatz (HVA):} 
    Inspired by the Quantum Approximate Optimization Algorithm (QAOA) and adiabatic quantum computation, HVA constructs the ansatz based on the problem Hamiltonian. 
    It is particularly effective for physics-inspired problems but introduces a hierarchical structure to manage complexity \cite{guo2024patterns}.
\end{itemize}

\subsection{Ansatzes for Quantum Machine Learning and Vision}
For high-dimensional data such as images, specialized ansatzes inspired by tensor networks and convolutional neural networks are preferred.

\begin{itemize}
    \item \textbf{Quantum Convolutional Neural Network (QCNN):} 
    QCNN adapts the structure of classical CNNs to the quantum domain. It consists of alternating convolutional layers (quasilocal unitary evolutions) and pooling layers (measurements or partial traces) to hierarchically reduce dimensionality while extracting features \cite{guo2024patterns}. 
    This structure effectively captures spatial correlations, making it suitable for image generation tasks.
    
    \item \textbf{Multiscale Entanglement Renormalization Ansatz (MERA):} 
    Originally designed for simulating quantum many-body systems, MERA represents quantum states using a hierarchical tensor network. It efficiently captures correlations at different length scales, which is analogous to the feature extraction process in deep learning \cite{guo2024patterns}.
\end{itemize}

\subsection{Optimal General Two-Qubit Ansatz (Vatan-Williams Decomposition)}
We adopt the optimal quantum circuit decomposition for general two-qubit interactions as proposed by Vatan and Williams \cite{Vatan_2004}. 
Specifically, we utilize the explicit circuit construction for the non-local interaction block $N(\alpha, \beta, \gamma)$, which is the core component of the ansatz.

\begin{figure}[h]
    \centering
    \begin{quantikz}
        \lstick{$q_0$} & \qw & \targ{} & \gate{R_z(\frac{\pi}{2}-2\gamma)} & \ctrl{1} & \qw & \targ{} & \gate{R_z(\frac{\pi}{2})} & \qw \\
        \lstick{$q_1$} & \gate{R_z(-\frac{\pi}{2})} & \ctrl{-1} & \gate{R_y(2\alpha-\frac{\pi}{2})} & \targ{} & \gate{R_y(\frac{\pi}{2}-2\beta)} & \ctrl{-1} & \qw & \qw
    \end{quantikz}
    \caption{The parameterized quantum circuit used in our Quantum Bottleneck. This structure implements the optimal decomposition for arbitrary two-qubit entangling gates, characterized by the parameters $\alpha, \beta, \gamma$. Note the alternating CNOT direction and the specific rotation angles derived from the canonical decomposition.}
    \label{fig:interaction_block}
\end{figure}
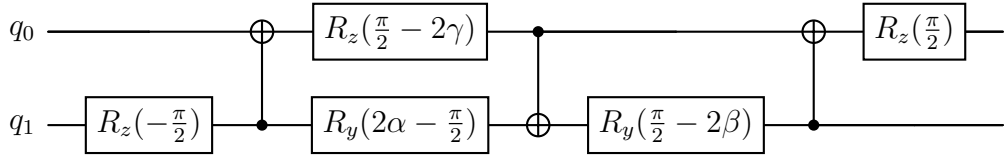

The circuit structure, depicted in Fig.~\ref{fig:interaction_block}, consists of three CNOT gates interleaved with parameterized single-qubit rotations. 
This configuration is capable of simulating the unitary operator $U = \exp(-i(\alpha \sigma_x \otimes \sigma_x + \beta \sigma_y \otimes \sigma_y + \gamma \sigma_z \otimes \sigma_z))$ up to local basis transformations.

The parameters $\alpha, \beta, \gamma$ serve as the learnable weights of the quantum diffusion model, allowing the network to capture complex correlations between features represented by qubits $q_0$ and $q_1$.

\subsection{Global Information Propagation: 1D Cluster State Mixing}
While the convolutional layer handles local correlations, effective generative modeling requires a mechanism to propagate information globally across the qubit register. 
To address this, we employ a \textit{Phase Mixing} layer inspired by 1D Cluster States (or Graph States) \cite{raussendorf2001one}.

The operation of this layer, denoted as $U_{\text{Mix}}$, proceeds in three stages designed to maximize information flow and adaptivity. 
First, the basis states are transformed into a uniform superposition via a layer of Hadamard gates ($H^{\otimes N}$), thereby initializing the system for parallel processing. 
Subsequently, we establish a linear entanglement backbone by applying a sequence of Controlled-Phase ($CZ$) gates between adjacent qubits. 
This creates a 1D cluster state structure that serves as a robust channel for propagating information across the register. 
Finally, to introduce spatial adaptivity, a layer of parameterized rotation gates $R_X(\phi_i)$ is applied to each qubit. 
Unlike uniform global strategies such as the Grover Mixer, these individual rotations allow the model to learn spatially specific weights, effectively distinguishing varying regions of the latent space, such as object boundaries versus backgrounds.

Mathematically, the mixing unitary is expressed as:
\begin{equation}
    U_{\text{Mix}}(\phi) = \left( \bigotimes_{j=1}^{N} R_X(\phi_j) \right) \left( \prod_{j=1}^{N-1} CZ_{(j, j+1)} \right) H^{\otimes N}
\end{equation}

\subsection{Global Feature Extraction via Hadamard Test} \label{sec:ancilla}
A unique component of our architecture is the inclusion of a global feature extractor using an ancilla qubit. 
As implemented in our circuit, we utilize a \textit{Hadamard Test} structure to measure the overlap between the current quantum state $\ket{\psi}$ and a parameterized unitary transformation $U(\theta)$.
The circuit prepares an ancilla in $\ket{+}$. 
A controlled-unitary operation $c\text{-}U(\theta)$ is applied, followed by a Hadamard gate and measurement on the ancilla. The expectation value provides:
\begin{equation}
    \langle Z_{ancilla} \rangle = \text{Re}(\langle \psi | U(\theta) | \psi \rangle)
\end{equation}
This scalar value represents a global property of the state (e.g., symmetry or macroscopic alignment) relative to the learned basis $U(\theta)$. In our hybrid model, this global feature is concatenated with the dense local features extracted by ANO, providing a multi-view representation to the classical decoder.

\section{Adaptive Non-Local Measurement}

In conventional Variational Quantum Circuits (VQCs), information extraction is typically limited to fixed local measurements, such as the expectation values of Pauli-$Z$ operators ($\langle \sigma_z^{(i)} \rangle$). 
This approach restricts the accessible information to local properties of individual qubits, failing to capture the rich correlations encoded in the entangled quantum state.

To fully leverage the high-dimensional Hilbert space of our quantum bottleneck, we implement the \textit{Adaptive Non-Local Observable (ANO)} framework \cite{lin2026quantumsuperresolutionadaptivenonlocal}. 
Instead of fixed operators, we employ a set of trainable Hermitian observables $\{H_k\}_{k=1}^{K}$ that act on the entire qubit register.

\subsection{Trainable Hermitian Construction}
For an $n$-qubit system with dimension $D=2^n$, a general observable is represented by a $D \times D$ Hermitian matrix. We parameterize the $k$-th observable $H_k(\phi_k)$ using a complex matrix $M_k(\phi_k) \in \mathbb{C}^{D \times D}$ with learnable weights $\phi_k$:
\begin{equation}
    H_k(\phi_k) = \frac{1}{2} \left( M_k(\phi_k) + M_k(\phi_k)^\dagger \right)
\end{equation}
This construction guarantees the Hermiticity ($H = H^\dagger$) required for valid quantum measurements while allowing the optimization process to explore the full manifold of observable operators.

\subsection{High-Dimensional Feature Extraction}
A key advantage of our approach is the ability to decouple the output feature dimension $K$ from the number of physical qubits $n$. By configuring $K \ge n$, we can extract an over-complete set of features from the latent quantum state $\rho = \ket{\psi(\theta)}\bra{\psi(\theta)}$. 
The output vector $\mathbf{y} \in \mathbb{R}^K$ fed into the classical decoder is given by:
\begin{equation}
    y_k = \bra{\psi(\theta)} H_k(\phi_k) \ket{\psi(\theta)} = \text{Tr}(\rho H_k(\phi_k)), \quad k=1, \dots, K
\end{equation}
By jointly training the circuit parameters $\theta$ and the measurement parameters $\phi$, the model learns to construct "quantum lenses" that focus on the specific sub-spaces containing the most relevant features for image reconstruction, such as high-frequency edge details.

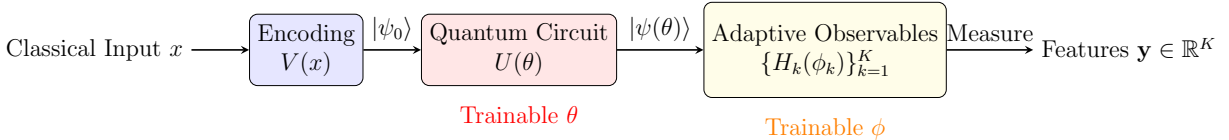
\begin{figure}[t] 
    \centering
    \resizebox{\textwidth}{!}{
        \begin{tikzpicture}
            \tikzstyle{block} = [draw, rectangle, minimum height=3em, minimum width=3em, align=center, rounded corners]
            \tikzstyle{arrow} = [thick,->,>=stealth]
            
            \node (input) {Classical Input $x$};
            \node[block, fill=blue!10, right=1cm of input] (encoder) {Encoding \\ $V(x)$};
            \node[block, fill=red!10, right=1cm of encoder] (ansatz) {Quantum Circuit \\ $U(\theta)$};
            
            \node[block, fill=yellow!10, right=1.5cm of ansatz, minimum height=4em] (ano) {Adaptive Observables \\ $\{H_k(\phi_k)\}_{k=1}^K$};
            
            \node[right=1.5cm of ano] (output) {Features $\mathbf{y} \in \mathbb{R}^K$};
            
            \draw[arrow] (input) -- (encoder);
            \draw[arrow] (encoder) -- node[above] {$\ket{\psi_0}$} (ansatz);
            \draw[arrow] (ansatz) -- node[above] {$\ket{\psi(\theta)}$} (ano);
            \draw[arrow] (ano) -- node[above] {Measure} (output);
            
            \node[below=0.2cm of ansatz, text=red] {Trainable $\theta$};
            \node[below=0.2cm of ano, text=orange] {Trainable $\phi$};
            
        \end{tikzpicture}
    }
    \caption{Schematic of the Quantum Bottleneck with Adaptive Non-Local Observables (ANO). 
    Unlike standard VQCs that output $N$ local measurements, our ANO framework employs a set of $K$ trainable Hermitian operators $\{H_k(\phi_k)\}$. 
    This allows the extraction of high-dimensional feature vectors ($K \ge N$) from the entangled quantum state $|\psi(\theta)\rangle$, effectively acting as a learnable quantum lens for super-resolution tasks.}
    \label{fig:ano_architecture}
\end{figure}

\section{Circuit Benchmarking and Evaluation}

Before integrating the Quantum Bottleneck into the full U-Net architecture, we quantitatively evaluated the intrinsic properties of our proposed ansatz. 
Following the framework established by Sim et al. \cite{Sim_2019}, we utilize two key descriptors: \textit{Expressibility} and \textit{Entangling Capability}. 
These metrics ensure that our circuit is sufficiently expressive to model complex latent distributions while effectively capturing non-local correlations required for the diffusion process.

\subsection{Expressibility via KL Divergence}

Expressibility measures the ability of a PQC to explore the Hilbert space. 
We quantify this by comparing the distribution of state fidelities generated by our ansatz against the distribution expected from an ensemble of Haar-random states (which represents the maximally expressive uniform distribution).

Let $F = |\braket{\psi_{\theta} | \psi_{\phi}}|^2$ be the fidelity between two states sampled from the PQC with random parameters $\theta, \phi$. 
The probability density function (PDF) of fidelities for Haar-random states in an $N$-dimensional Hilbert space ($N=2^n$) is given by $P_{\text{Haar}}(F) = (N-1)(1-F)^{N-2}$.

The expressibility $E$ is defined as the Kullback-Leibler (KL) divergence between the PQC's fidelity distribution $P_{\text{PQC}}(F)$ and the analytical Haar distribution $P_{\text{Haar}}(F)$:
\begin{equation}
    E = D_{KL}(P_{\text{PQC}} || P_{\text{Haar}}) = \int_0^1 P_{\text{PQC}}(F) \ln \left( \frac{P_{\text{PQC}}(F)}{P_{\text{Haar}}(F)} \right) dF
\end{equation}
A lower value of $E$ indicates that the ansatz can uniformly explore the Hilbert space, approaching the statistical properties of random states. 
In our experiments, we approximate this integral using a discretized histogram of sampled fidelities.

\subsection{Entangling Capability via Meyer-Wallach Measure}

To verify the circuit's ability to generate multi-partite entanglement—a crucial feature for capturing global dependencies in image data—we employ the Meyer-Wallach (MW) measure \cite{Meyer_2002} $Q$. 
For a given state $\ket{\psi}$, the MW measure is defined as the average purity of the single-qubit reduced density matrices:
\begin{equation}
    Q(\ket{\psi}) = \frac{4}{n} \sum_{k=1}^{n} D(\iota_k(\ket{\psi})) = \frac{4}{n} \sum_{k=1}^{n} \frac{1}{2} \left( 1 - \text{Tr}(\rho_k^2) \right)
\end{equation}
where $\rho_k = \text{Tr}_{\neg k}(\ket{\psi}\bra{\psi})$ is the reduced density matrix of the $k$-th qubit. The value $Q$ ranges from 0 (product states, unentangled) to 1 (maximally entangled, e.g., GHZ states).

We estimate the \textit{Entangling Capability} \cite{Sim_2019} of our ansatz by averaging $Q$ over an ensemble of states sampled with uniformly random parameters:
\begin{equation}
    \bar{Q} = \frac{1}{S} \sum_{i=1}^{S} Q(\ket{\psi(\theta_i)})
\end{equation}
A high $\bar{Q}$ value confirms that our \textit{ConvUnit} and \textit{PhaseMixing} layers successfully distribute information across the qubit register, facilitating the "global mixing" required for high-resolution image synthesis.

\subsection{Visualizing the State Space}
\begin{figure}[H]
    \centering
    \includegraphics[width=0.85\linewidth]{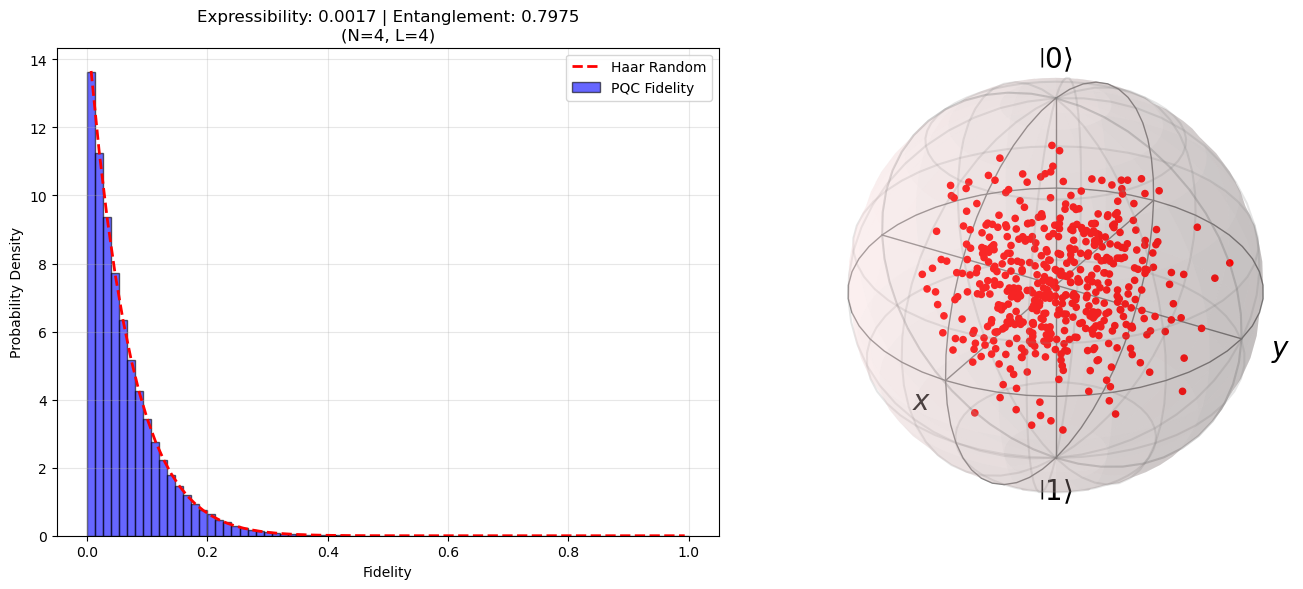}
    \caption{Visualization of the single-qubit reduced state distribution on the Bloch sphere generated by our ANO-based ansatz. 
    Each red point represents the state of the first qubit traced out from the multi-qubit system, sampled over random circuit parameters. 
    (Left) Points distributed strictly on the surface indicate product states with no entanglement. 
    (Right) Points filling the interior volume demonstrate the generation of strong multi-partite entanglement, as the reduced state becomes mixed due to correlations with other qubits. 
    This volumetric coverage visually corroborates the high Meyer-Wallach measure and Expressibility scores.}
    \label{fig:bloch_dist}
\end{figure}

In addition to quantitative metrics, we visualize the distribution of single-qubit reduced states on the Bloch sphere. 
As shown in Fig.~\ref{fig:bloch_dist}, the concentration of states within the interior of the Bloch sphere (as opposed to the surface) serves as visual evidence of entanglement, consistent with high Meyer-Wallach values.

\section{Architecture}
\begin{figure}[htbp]
  \centering
  \includegraphics[width=\linewidth]{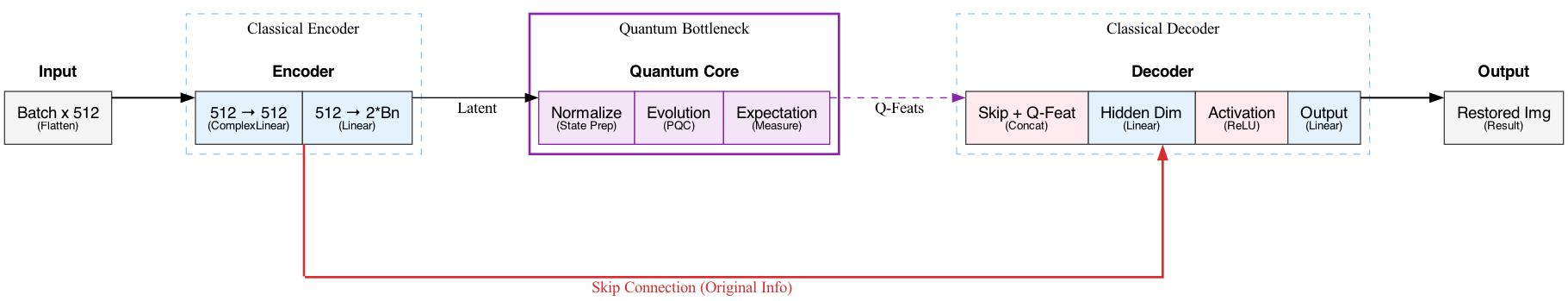}
  \caption{Schematic of the Hybrid Quantum-Classical U-Net Architecture. The model integrates a quantum bottleneck for feature extraction and employs a skip connection to preserve original semantic information during the reverse diffusion process.}
  \label{fig:architecture}
\end{figure}

The proposed model, illustrated in Figure \ref{fig:architecture}, adopts a hybrid quantum-classical U-Net architecture inspired by recent advances in quantum generative diffusion models \cite{cacioppo2023quantumdiffusionmodels, qdm}. 
The architecture is designed to leverage the expressivity of quantum circuits while maintaining the reconstruction capability of classical networks. 
It consists of three distinct modules: a classical encoder, a quantum bottleneck (core), and a classical decoder.

First, the Classical Encoder transforms the input data into a lower-dimensional latent representation suitable for the quantum interface. 
Due to the simulation overhead of quantum circuits, the original MNIST images are downsampled to a resolution of $16 \times 16$, resulting in a flattened input vector of dimension $2^8=256$.
To seamlessly interface with the quantum layer, we utilize complex-valued linear layers (\texttt{ComplexLinear}) that preserve phase information necessary for quantum state preparation.

Second, the Quantum Bottleneck acts as the core generative component. 
We employ \textit{amplitude encoding} to efficiently map the classical latent vector $\mathbf{z}$ onto an $n=4$ qubit quantum state $\ket{\psi}$ by normalizing the complex vector. 
This state is evolved by a Parameterized Quantum Circuit (PQC), $U(\theta)$, which learns the reverse diffusion dynamics. 
Crucially, our bottleneck employs a dual-path feature extraction strategy to capture both local and global information.
The evolved state is measured using Adaptive Non-Local Observables (ANO) to extract high-dimensional dense features.
Simultaneously, we integrate an Ancilla-based Hadamard Test module, which utilizes an auxiliary qubit to measure the global overlap between the state and a parameterized unitary. 
This global scalar feature is concatenated with the dense ANO features, providing a rich, multi-view representation of the quantum latent space.

Finally, the Classical Decoder reconstructs the denoised sample from the aggregated quantum features. 
A crucial design element is the inclusion of a Skip Connection (red path in Figure \ref{fig:architecture}). 
This connection concatenates the original flattened input vector directly with the quantum features in the decoder. 
This mechanism is essential for mitigating the "bottleneck" problem inherent in low-qubit quantum circuits (compressing 256 dimensions to 4 qubits) and prevents the model from converging to trivial identity mappings by preserving the original signal structure.

\section{Experimental Results}
In this section, we present the evaluation of the proposed Hybrid Quantum-Classical U-Net. 
The model was trained on the full MNIST dataset (digits 0 through 9) to assess its ability to learn multi-modal distributions. 
Due to the computational overhead associated with simulating parameterized quantum circuits (PQC), the original MNIST images were downsampled to a resolution of $16 \times 16$ for all experiments.
We benchmark our results against the limitations observed in prior quantum diffusion studies \cite{cacioppo2023quantumdiffusionmodels}, particularly regarding multi-class scalability.

\subsection{Generative Process Dynamics: From Chaos to Structure}
\begin{figure}[htbp]
    \centering
    \includegraphics[width=1.0\textwidth]{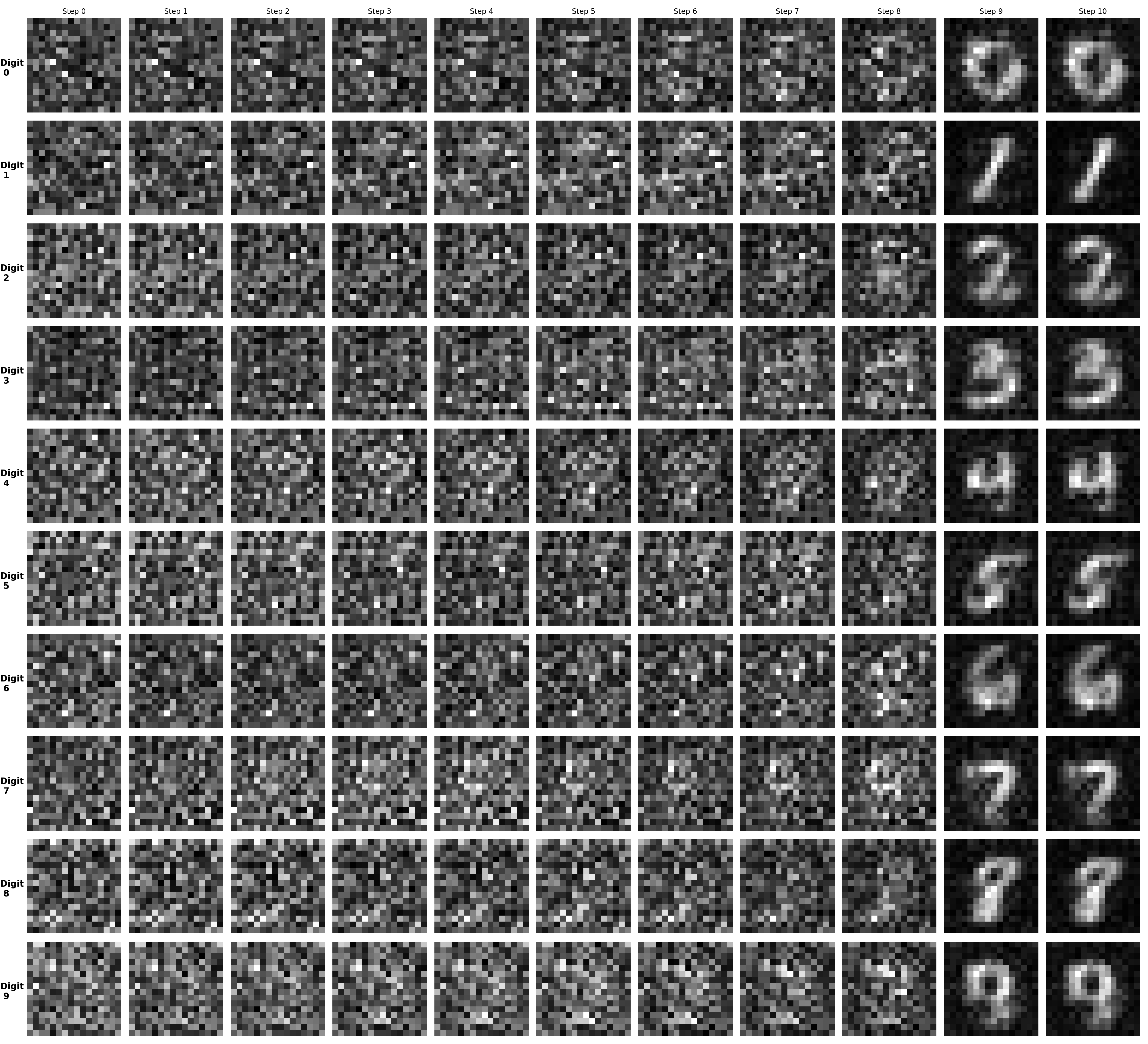}
    \caption{\textbf{Dynamics of Reverse Diffusion.} The columns represent discrete time steps from pure noise (Step 0) to the final generated sample (Step 10). The successful reconstruction of distinct digit topologies from random noise validates that the quantum model has learned the underlying probability flow of the multi-modal distribution.}
    \label{fig:generation}
\end{figure}

The core capability of a diffusion model lies not merely in the final image quality, but in learning the precise \textit{reverse diffusion trajectory} that maps a Gaussian noise distribution back to the data manifold. We visualized this step-by-step denoising process in Figure \ref{fig:generation}.

As illustrated, the model successfully initiates from pure Gaussian noise ($t=T$) and progressively suppresses entropy to recover structural patterns. 
The transition from abstract noise to recognizable digits confirms that the parameterized quantum circuit (PQC) has effectively learned the \textit{reverse transition kernel} $p_\theta(x_{t-1}|x_t)$. 
Crucially, the distinct formation of digits at the final step ($t=0$) demonstrates that the Hybrid U-Net preserves semantic information throughout the denoising chain. 
Despite the information bottleneck imposed by the 4-qubit quantum core, the model retains the global topology required to distinguish between complex classes (e.g., distinguishing a '3' from an '8'), a feat that goes beyond simple memorization and validates the model's understanding of the underlying probability flow.

\subsection{Multi-Class Generation and Scalability}
A significant challenge in Quantum Machine Learning (QML) has been scaling generative models to multi-class datasets. 
Previous works demonstrated generation capabilities primarily on binary subsets or required heavy classical post-processing for larger sets.

\begin{figure}[htbp]
    \centering
    \includegraphics[width=0.8\textwidth]{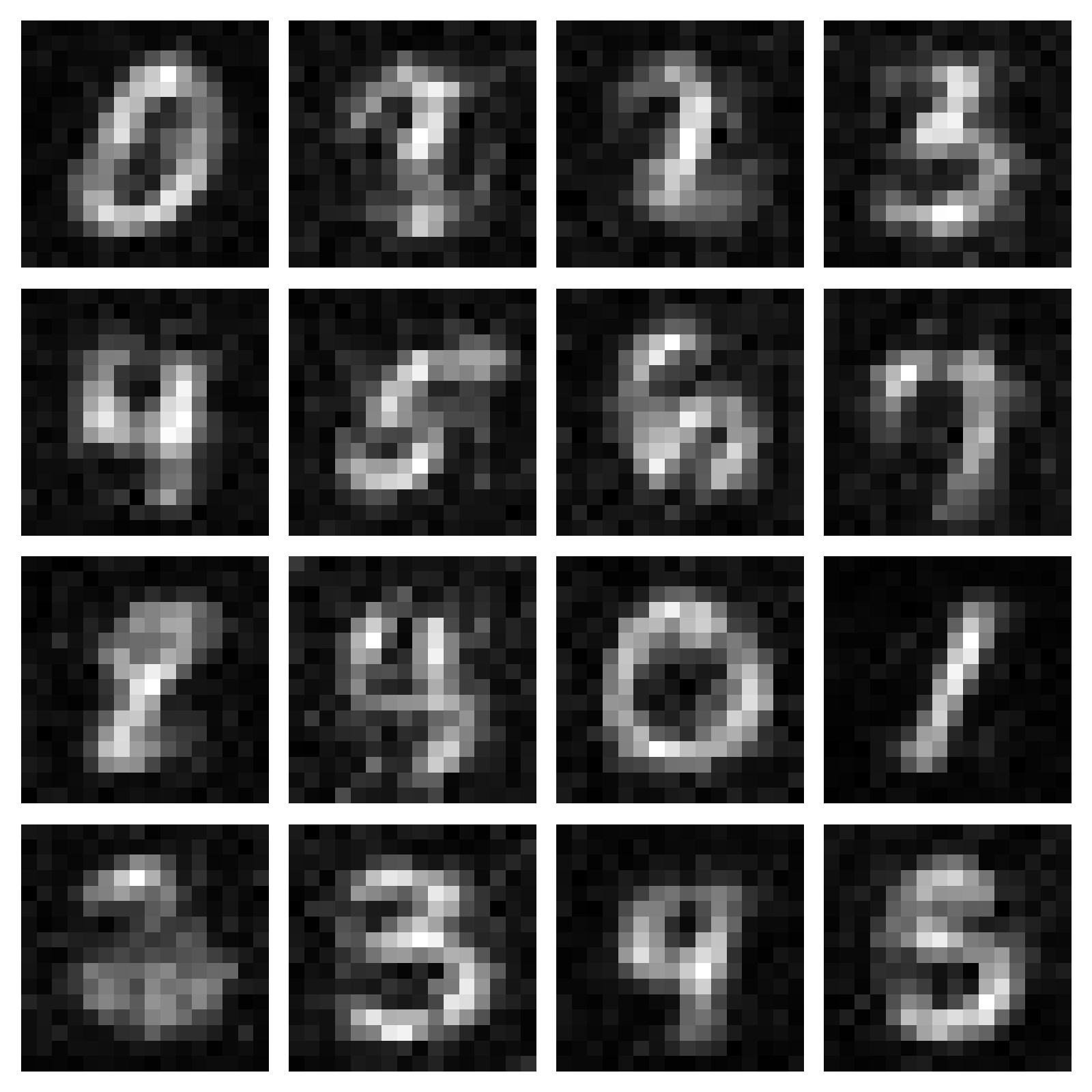}
    \caption{\textbf{Multi-Class Generation Results (Digits 0-9).} The grid displays random samples generated by the trained model. In contrast to prior studies limited to binary classes, our Hybrid Quantum U-Net successfully generates recognizable digits across all ten classes, demonstrating feasibility for more complex multi-modal tasks.}
    \label{fig:sampling}
\end{figure}

Figure \ref{fig:sampling} presents a batch of uncurated samples generated by our model trained on all ten digit classes. 
The results show that the model generates distinct samples for all digits (0-9), mitigating the mode collapse often observed in pure quantum baselines. This indicates that the dual mechanism of ANO (local features) and Ancilla-based measurement (global features) allows the quantum bottleneck to capture sufficiently rich correlations to distinguish between multiple modes. While the generated images exhibit some blurring—a known trade-off in bottlenecked hybrid architectures and resolution downsampling—the structural coherence is maintained. Key morphological features are recognizable, suggesting that the expressibility of our ansatz is sufficient to encode the semantic information of the full MNIST dataset.

\subsection{Quantitative Evaluation}
To strictly assess the generative quality, we evaluated the Fréchet Inception Distance (FID). 
This metric measures the distance between the distribution of generated images and real images in the feature space of a pre-trained InceptionV3 network \cite{heusel2018ganstrainedtimescaleupdate, szegedy2015rethinkinginceptionarchitecturecomputer}.

Our optimized model achieved an FID score of approximately 150. 
While numerical metrics in quantum generative models are often penalized by resolution constraints, this score represents a significant achievement given the \textit{extreme information compression} imposed by the quantum circuit. 
The core generative process relies on a bottleneck of only $n=4$ qubits (Hilbert space dimension $2^4=16$), forcing the model to reconstruct $256$-dimensional data from a highly compressed quantum latent space.

\begin{figure}[htbp]
    \centering
    \includegraphics[width=0.8\textwidth]{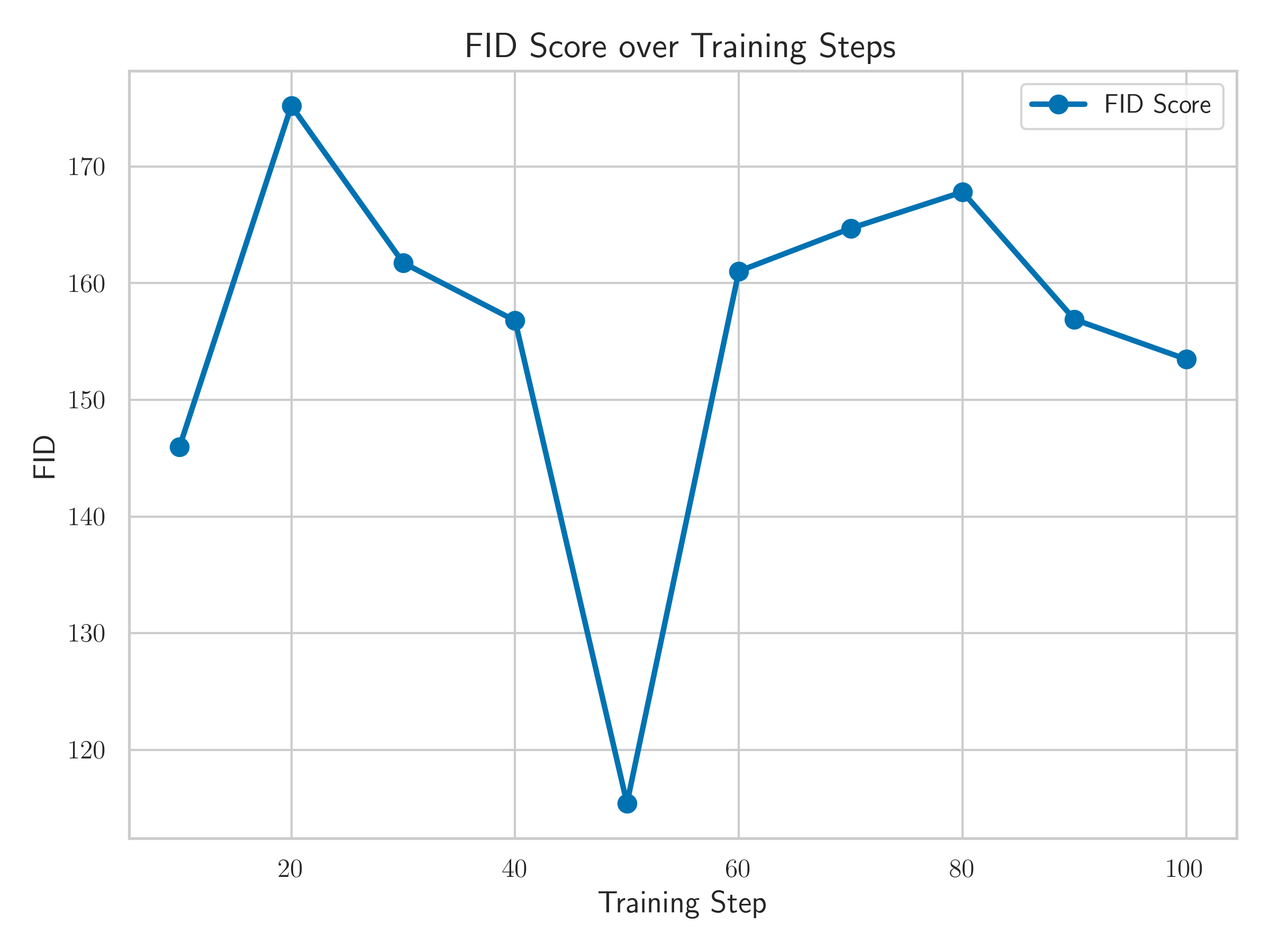}
    \caption{\textbf{Evolution of Fréchet Inception Distance (FID) during training.} The graph demonstrates a rapid improvement in generative quality during the initial epochs, stabilizing around a score of 110. This convergence trajectory confirms that the Hybrid Quantum U-Net effectively learns the data distribution despite the highly constrained qubit bottleneck.}
    \label{fig:fid_plot}
\end{figure}

To contextualize this performance, we established a "Real-to-Real" baseline FID by comparing two subsets of real MNIST images that underwent the same $16 \times 16$ downsampling and upscaling process. 
This baseline yielded a score of 8.79. The narrowed gap between the baseline and our model highlights the effectiveness of the proposed hybrid architecture. 
It suggests that the high FID is largely an artifact of the resolution mismatch (which sets the floor at $\approx 8.8$) rather than a failure of the generative process. The fact that the model achieves a score of 110 despite the \textbf{16:1 compression ratio} (256 classical dimensions to 16 quantum states) confirms that the Adaptive Non-Local Observables (ANO) successfully extract the essential semantic features required for high-fidelity reconstruction.

\section{Conclusion}
In this study, we proposed and validated a Hybrid Quantum-Classical U-Net architecture designed to overcome the scalability and expressibility limitations of quantum generative models. 
By integrating Adaptive Non-Local Observables (ANO) and an Ancilla-based global feature extractor within a classical encoder-decoder framework, we successfully demonstrated the generation of complex, multi-modal distributions using the full MNIST dataset (digits 0-9).

The experimental results yield several critical insights regarding the potential of hybrid generative models. 
First, the model effectively mitigates mode collapse, a persistent challenge in prior quantum diffusion studies, by generating distinct and recognizable images for all ten digit classes. 
This confirms that the proposed hybrid architecture, specifically the dual-path quantum bottleneck, captures sufficient semantic information to reconstruct high-dimensional data manifolds. 
Second, the study verifies the structural efficiency of compressing classical features before mapping them to the quantum Hilbert space. 
This approach leverages quantum expressivity while circumventing the exponential resource overhead associated with direct high-qubit encoding. 
Furthermore, we identified that skip connections are indispensable for preserving structural integrity. 
By facilitating the direct flow of high-resolution semantic information to the decoder, these connections compensate for the information loss inherent in the low-qubit bottleneck, ensuring that the generated images retain sharp morphological features.

Despite these achievements, we observed that the Fréchet Inception Distance (FID) score exhibits oscillations ranging between 110 and 160 during the convergence phase, rather than settling at a global minimum. 
This instability suggests that the optimization landscape of the parameterized quantum circuit is highly non-convex and potentially hindered by barren plateaus, a common hurdle in training deep quantum circuits. 
For future work, it would be valuable to investigate whether advanced optimization strategies, such as Quantum Natural Gradient Descent (QNGD) to adjust gradient updates according to the quantum information geometry (Fubini-Study metric \cite{cheng2013quantumgeometrictensorfubinistudy}), could mitigate this oscillation by accounting for the quantum information geometry. 
However, further research is required to determine the precise cause of the instability and to explore diverse stabilization techniques. 
Additionally, extending the current simulation-based results by deploying the inference pipeline on real NISQ hardware remains a key objective to validate the practical robustness of our adaptive measurement strategy in noisy environments.

\section*{Acknowledgments}
This research was supported by the \textit{QAMP 2025} program. \textit{Qiskit Advocate Mentorship Program (QAMP)} is a program focused on bringing new contributors into Qiskit open source software development where Qiskit advocates work on a 3-month project under the guidance of mentors. 
It is an initiative within the \href{https://www.ibm.com/quantum/qiskit#advocates}{Qiskit advocate program} designed to support growth and collaboration within our vibrant community.
Access the QAMP information deck \href{https://github.com/qiskit-advocate/qamp-2025/blob/main/QAMP_call.pdf}{here}. 
We thank the mentors and organizers for their guidance and computing resources provided for this project.

\section*{Code Availability}
To support the reproducibility of this research, the source code for the Hybrid Quantum-Classical U-Net implementation is made publicly available.
This includes the Parameterized Quantum Circuit (PQC) designs based on PennyLane, the Adaptive Non-Local Observables (ANO) framework, and the training scripts used for the MNIST experiments. 
The repository can be accessed at: \href{https://github.com/dolf3131/Enhancing-Quantum-Diffusion-Models-for-Complex-Image-Generation}{github}.

\printbibliography

\end{document}